\documentclass[a4paper,aps,pra,reprint]{revtex4-1}
\usepackage{amsmath}
\usepackage{graphicx}
\usepackage{amssymb}
\usepackage{color}
\usepackage{bbold} 

\newcommand{\ff}[0]{\omega}
\newcommand{\Ls}[0]{\mathcal{L}}
\newcommand{\Ms}[0]{\mathcal{M}}
\newcommand{\Us}[0]{\mathcal{U}}
\newcommand{\refq}[1]{(\ref{#1})} 
\newcommand{\eqexp}[1]{\left\langle #1 \right\rangle_{\rm B}^{\rm eq}}
\newcommand{\expe}[1]{\left \langle #1 \right \rangle_{\rm B}}

\begin{document}
\title{Dynamics of a quantum two-state system in a linearly driven quantum bath}

\date{\today}
\author{J. Reichert,$^{1,2}$ P. Nalbach,$^3$ and M. Thorwart$^{1,2}$}
\affiliation{$^1$I.\ Institut f\"ur Theoretische Physik, Universit\"at Hamburg,
Jungiusstra{\ss}e 9, 20355 Hamburg, Germany \\
$^2$The Hamburg Centre for Ultrafast
Imaging, Luruper Chaussee 149, 22761 Hamburg, Germany \\
$^3$Westf\"alische Hochschule, M\"unsterstra{\ss}e 265, 46397 Bocholt, Germany}

\begin{abstract}
When an open quantum system is driven by an external time-dependent force, 
the coupling of the driving to the central system is usually included whereas 
the impact of the driving field on the bath is neglected. We investigate 
the effect of a quantum bath of linearly driven harmonic oscillators on the 
relaxation dynamics of a quantum two-level system which itself is not directly 
driven. In particular, we calculate the 
frequency-dependent response of the system when the bath is
subject to a Dirac and a Gaussian driving-pulse. We show 
that a time-retarded effective force on the system is induced by the driven 
bath which depends on the full history of the perturbation and the spectral 
characteristics of the underlying bath. In particular, when a structured Ohmic 
bath with a pronounced Lorentzian peak is considered, the dynamical response of 
the system to a driven bath is qualitatively different as compared to the 
undriven bath. Specifically, additional resonances appear which can be directly 
associated to a Jaynes-Cummings-like effective energy spectrum. 
\end{abstract}

\maketitle

\section{Introduction}

The effect of environmental fluctuations on the dynamics of quantum systems
has been a longstanding focus of quantum statistical physics. 
Indeed, the modern fields of quantum dissipation and open quantum systems 
treat a physical problem by separating it into an identifiable ``system'', which 
encompasses a few controllable degrees of freedom, and an environmental
``bath'' (or reservoir) \cite{Weiss_2012}, which consists of infinitely many 
degrees of freedom and exerts fluctuating forces on the central system. 
In many physical situations, the spectral statistics of the (classical or 
quantum) fluctuations is Gaussian, such that the underlying physical 
model of a harmonic bath is adequate. Thereby, infinitely many harmonic 
oscillators are used in conjunction with a bi-linear system-bath coupling. After 
integrating over the harmonic bath degrees of freedom, a reduced density 
operator of the system is constructed whose effective non-unitary 
time-dependence is studied. 

A minimal model system which allows to study the role of dissipative 
fluctuations on the transition dynamics between two quantum mechanical states 
is the spin-boson model. It describes a quantum two-level system (TLS) coupled 
to a bath of harmonic oscillators \cite{LeggettPaper_1987} 
and has been used for the analysis of such diverse physical 
phenomena as tunneling of defects in low-temperature amorphous materials  
\cite{glass1,glass2,NalbachGlassesPaper_2004,glasses}, the role of the solvent
on electron transfer in 
chemical reactions \cite{Nitzan_2006}, energy transfer in biomolecular 
photoactive complexes \cite{AmerongenBook}, 
or the analysis of decoherence and relaxation properties of 
solid-state qubit devices realized in single-charge and spin quantum dots 
\cite{qdotreview} and superconducting quantum interference devices
\cite{Schoen01}.

A useful tool to investigate quantum systems is through application of a
time-dependent external field 
\cite{Grifoni98,Weiss_2012,driventunnel1,driventunnel2}. 
The impact of such time-dependent driving is usually included by 
way of a direct coupling of the external field to the central system of 
interest, while the impact of the time-dependent driving on the environment is 
not included. It was recently realized, however, that ancillary 
driving of the reservoir itself is unavoidable in principle on the nanoscale in 
many physical applications \cite{GrabertPaper_2016_drivenbaths}. 
In fact, as it was recently shown for two model systems, the system of interest 
becomes subject to an additional bath-induced force component, 
if an external time-dependent drive couples to the bath as well. 
For instance, the exact solution of the polarizability of a test 
molecule immersed in water which is also subject to external driving, reveals
\cite{GrabertPaper_2016_drivenbaths} that the frequency-dependent response is
increased by about $30\%$ as compared to the case when bath-driving 
is not considered. Moreover, the frequency-dependent response 
of a semiconducting nanocrystal placed in the vicinity of a metallic
nanoparticle and both immersed in a solvent, was shown to be qualitatively
altered when bath-driving is included \cite{GrabertPaper_2016_drivenbaths}.
Furthermore, the impact of coupling external driving to the 
environmental modes of driven superconducting tunnel junctions was shown 
to yield significant contributions 
\cite{GrabertPaper_2015_blockade,FreyPaper_2016}. 

In this work, we study the way in which
time-dependent driving of the harmonic modes of a quantum bath affects the 
relaxation dynamics of a quantum two-level system. This constitutes a 
generalization of the spin-boson model in which an external time-dependent 
driving force is coupled linearly to the bath. For simplicity, we do not 
consider an additional direct coupling of the external time-dependent force to the system itself, but focus 
our attention on the impact of the driven bath. We show 
explicitly that the driven bath generates a time-retarded effective force which acts on 
the two-level system. We address the relaxation dynamics in the regime of 
weak system-bath coupling such that the effective dynamics of the central 
quantum system can be described in terms of a quantum master equation with time-dependent 
rate coefficients. Specifically, we 
apply a suitable adiabatic Born-Markov approximation 
\cite{NalbachPaper_2013_am,NalbachPaper_2014_am,WuergerPaper_1997, 
NalbachDiss_1998, Weiss_2012} which is valid for not too fast driving. We 
consider two types of bath spectral densities, the simple structureless Ohmic 
bath and a structured Ohmic bath which contains a single pronounced harmonic 
mode. The latter is known to be equivalent to a cavity QED setup 
\cite{ThorwartPaper_2004_structured}. We calculate the response of the 
dissipative quantum two-level system to 
time-dependent bath-driving with a $\delta$-shaped as well as a
Gaussian-shaped driving pulse. The effective bath-induced force is given in
exact form. We show that the response of the two-level system to the driven
bath is noticeably altered. For the unstructured Ohmic bath, the resonant
response of the quantum two-level system decreases in a driven bath as compared 
to the undriven case. A qualitatively different response
arises when the structured Ohmic bath is driven. Additional resonant peaks
appear in the response of the system when the external 
drive matches resonances related to environmental modes. The 
present formulation within the general context of quantum dissipation could 
potentially open up novel pathways to control the dynamics of 
quantum two-level systems by manipulating their environment through  
time-dependent control fields.

\section{Model Hamiltonian}

We consider a model of a quantum mechanical two-level system which is coupled
to a linearly driven quantum bath. The associated total time-dependent 
Hamiltonian 
\begin{equation}
\label{SB-1}
 H(t) = H_{\rm S} + H_{\rm SB} + H_{\rm B} + H_{\rm IB}(t)
\end{equation}
is the sum of a system Hamiltonian $H_{\rm S}$ 
and a bath Hamiltonian $H_{\rm B}$, along with a part $H_{\rm SB}$ that
describes the system-bath coupling. The new
term $H_{\rm IB}(t)$ represents the effect of external
time-dependent driving on the bath.

To be specific, we consider in this work a symmetric quantum
mechanical two-state system ($\hbar = 1$ and $\sigma_{i=x,z}$ denote the Pauli
matrices) with 
\begin{equation}
\label{sysham}
H_{\rm S}  = \frac{\Delta}{2}\sigma_{x} \text{,}
\end{equation}
which couples to a bath of harmonic oscillators 
\begin{equation}
\label{bathham}
H_{B} = \sum_{j}^{N} \omega_{j} \left(b^{\dagger}_{j} b_{j} + \frac{1}{2}\right)
\text{,}
\end{equation}
via
\begin{equation}
\label{sysbathham}
H_{\rm SB} = - \frac{\sigma_{z}}{2} \sum_{j}^{N} c_j \left(b_j +
b^{\dagger}_j\right) 
\end{equation}
with coupling constants $c_j$. Here, $b_{j}$ and $b^{\dagger}_{j}$ denote the
corresponding annihilation and creation operators of the $j$-th bath mode. 
The new term describes the coupling of the bath to an external, classical force 
$F(t)$ and is included as
\begin{equation}
\label{SB-3}
 H_{\rm IB}(t) = -\frac{F(t)}{2} \sum_{j}^{N} d_j \left( b_j +  b^{\dagger}_j
\right) \, ,
\end{equation}
where the $d_j$ denote the associated coupling constants. 
The driving of each bath mode is assumed to be of dipolar type, coupling
to the displacement of the oscillators. This linear (or additive) form of the
coupling 
does not modify the mean square displacements of the oscillators and, thus,
does not alter the (equilibrium) temperature of the bath which is fixed at the
initial time (see below). This would be different if the external force coupled
parametrically, i.e., $H_{\rm IB}^{\rm para}(t) = -\frac{F(t)}{2} \sum_{j}^{N}
d_j b_j^\dagger b_j$.

As usual \cite{Weiss_2012}, we characterize the bath by the spectral density 
\begin{equation}
\label{SB-4}
J(\ff) = \pi \sum_{j}^{N} c^{2}_j \delta(\ff - \ff_j)\text{,}
\end{equation}
with specific forms of $J(\ff)$ given below.

\begin{figure}[t!]
\centering
\includegraphics[width=9cm]{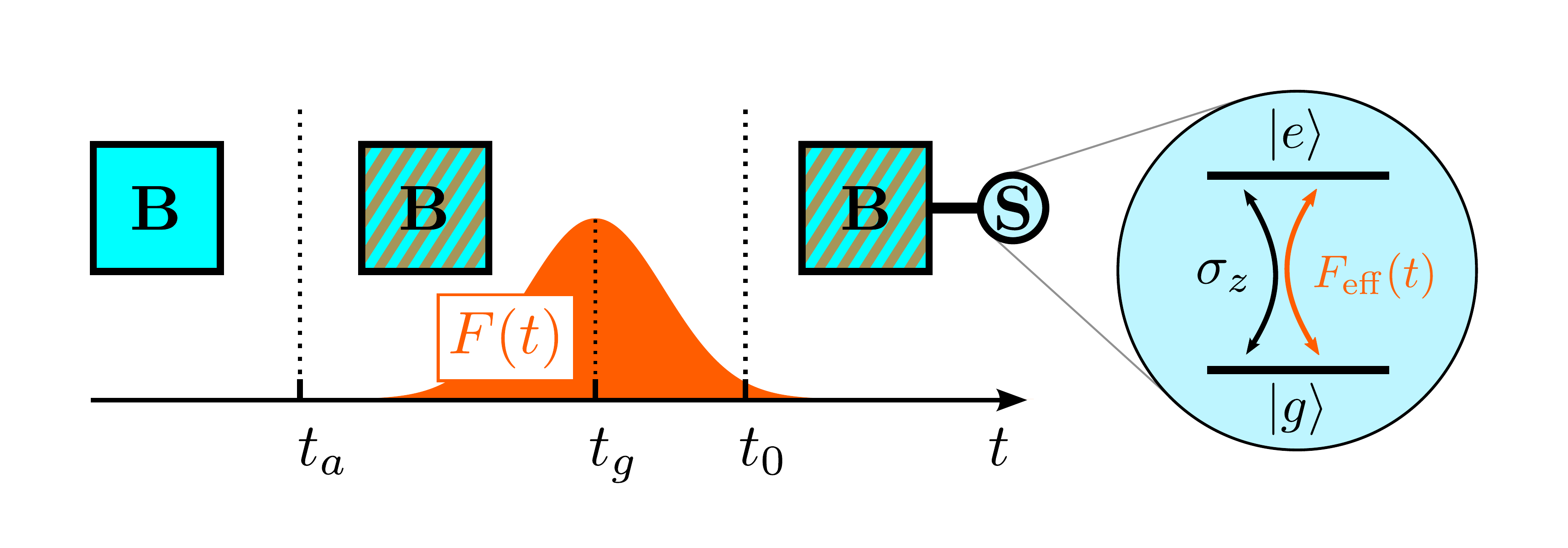}
\caption{\label{fig-1} General setup of a pulse-shaped bath drive. The bath (B)
is in equilibrium until time $t_{a}$, when the bath driving force $F(t)$ 
(orange and shown as a generic Gaussian pulse) is activated. Subsequently, the 
bath is driven (orange stripes), with the
perturbation centered at some time $t_{g}$. At time $t_{0}$ 
the system (S) is coupled to the driven bath. We consider a 
two-level system with the ground (g) and excited
state (e) coupled to the harmonic bath. Bath driving leads to an additional 
effective force $F_{\rm eff}(t)$, as shown in this work. 
}
\end{figure}

The time dependence of the bath Hamiltonian requires some attention in view of 
the initial condition for the dissipative dynamics. The most convenient choice
is factorizing initial conditions. In this case, the system is assumed 
to be initially decoupled from the bath and the coupling is switched on 
instantaneously at time $t_0$ \cite{Weiss_2012}. For the time-dependent 
bath-driving of Eq. \refq{SB-3},
we consider pulse-shaped driving, in particular a $\delta$-shaped
and a Gaussian pulse, starting at time $t_a$. 
Figure \ref{fig-1} shows the scheme which we follow throughout this work. 
We assume the bath to be in thermal equilibrium until time $t_{a}$. 
Then, the density matrix of the bath is given by $\rho_{\rm B}(t_{a}) = 
\rho_{\rm B}^{\rm eq} = e^{-\beta H_{\rm B}}/\mathcal{Z}$ at the given 
temperature $T=1/\beta$ with $k_B=1$
($\mathcal{Z}$ is the equilibrium partition of the decoupled bath). At 
$t=t_{a}$, the action of the pulse on the bath is turned on and the
interaction $F(t)$ in Eq.\ \refq{SB-3} becomes nonzero. Subsequently, the 
bath evolves under the combined time evolution operator defined by
\begin{equation}
\label{hbatheff}
 H_{\rm B}^{\rm eff}(t) = H_{\text{B}} + H_{\rm IB}(t)\text{.}
\end{equation}

In addition, we consider the system-bath coupling $H_{\rm SB}$ to be active 
for times $t > t_{0}$ onwards.

\section{Driven bath dynamics}

Due to the additivity of the external driving, it is convenient to address 
the Heisenberg operators $\tilde{b}_{j}(t)$ of the bath. Therefore, we consider
the time evolution of each bath operator $b_{j}$ under the driven
bath Hamiltonian of Eq.\ \refq{hbatheff}. The
Heisenberg operators $\tilde{b}_{j}(t)$ are found to be of the form 
\cite{Merzbacher_1998}
\begin{equation}
\label{Force-1}
  \tilde{b}_{j}(t) = \tilde{b}^{0}_{j}(t) + \frac{1}{2}K_j(t,t_{a})
\end{equation}
with the Heisenberg operator under force-free time evolution
\begin{equation}
 \label{Force-10}
 \tilde{b}^{0}_{j}(t) = b_{j}e^{-i\omega_{j} t}\, ,
\end{equation}
and the driving-induced term
\begin{equation}
 \label{Force-2}
 K_j(t,t_{a}) = i\int_{t_{a}}^{t} dt' e^{i\ff_j (t'-t)} d_j F(t'). 
\end{equation}
The corresponding equation for $\tilde{b}^{\dagger}_{j}(t)$ can be obtained by
standard Hermitian conjugation. 

\subsection{Effective force}
\label{sub:force}

Since each bath oscillator is not statically displaced, it follows that 
$\tilde{b}^{0}_{j}(t)$ has zero average at equilibrium, i.e., $\langle
\tilde{b}^{0}_{j}(t) \rangle_{\rm B}^{\rm eq} = 0$. However, due to Eq.\ 
\refq{Force-2}, linear bath driving induces a nonzero contribution, such that  
$\langle \tilde{b}_{j}(t) \rangle_{\rm B}^{\rm eq} = K_j(t,t_{a})/2$.
This implies
\begin{equation}
\label{Force-3}
\begin{split}
\langle H_{\rm SB} & \rangle_{\rm B} (t) = -\frac{\sigma_{z}}{2} 
\sum_{j}^{N} c_j \eqexp{\tilde{x}_j(t)} \\
& = -\frac{\sigma_{z}}{2} {\rm Re}\left[\sum_{j}^{N} c_j K_{j}(t,t_{a}) \right]
\equiv 
\frac{\sigma_{z}}{2}  F_{\text{eff}}(t)\text{,}
\end{split}
\end{equation}
where we inserted the dimensionless (Heisenberg) position operator 
$\tilde{x}_j(t) = \tilde{b}_{j}(t) + \tilde{b}^{\dagger}_{j}(t)$. This defines 
the effective force $F_{\text{eff}}(t)$ which can be
formulated in a convenient way by introducing an additional spectral density
\begin{equation}
\label{Force-4}
\bar{J}(\ff) = \pi \sum_{j}^{N} d_j c_j \delta(\ff - \ff_j)\text{.}
\end{equation}
It incorporates the system-bath coupling constants $c_j$ as well as
the coupling constants of the external driving to the bath $d_j$. With this,
the continuum limit of an infinitely dense spectrum of environmental modes can
be performed. Then, the effective force follows as 
\begin{equation}
\label{Force-5}
F_{\text{eff}}(t) = {\rm Im}\left[\frac{1}{\pi} \int_{0}^{\infty}d\ff
\bar{J}(\ff) \int_{t_{a}}^{t} dt' F(t') e^{i\ff (t'-t)} \right]\text{.}
\end{equation}
It should be emphasized that the effective force is time-dependent and
only nonzero for times $t > t_0$ as it depends on the system-bath
couplings $c_j$.

\subsection{Fluctuations}
\label{sub:fluct}

As the driving-induced term in Eq.\ \refq{Force-2} is proportional to the
identity operator, a simple shift 
of the Heisenberg operator $\tilde{b}_{j}(t)$ by the average $\langle
\tilde{b}_{j}(t) \rangle_{\rm B}^{\rm eq}$ allows us to 
recover an effective undriven time evolution. Accordingly, the shifted (Heisenberg)
position operator fulfills 
$\tilde{x}_{j}^{\text{eff}}(t) = \tilde{x}_j(t) - \eqexp{\tilde{x}_j (t)} = 
\tilde{x}^{0}_j(t)$. Consequently, the bath autocorrelation function 
$B_{\text{C}}(t,s)$ and the bath response
function $B_{\text{R}}(t,s)$ remain unchanged compared to their  
equilibrium form. In particular, we have that 
\begin{equation}
\label{Force-6}
\begin{split}
 B_{\text{C}}(t,s) & = \eqexp{\sum_{j,j'}^{N}\frac{c_{j} c_{j'}}{2}
\left\{\tilde{x}^{\text{eff}}_{j}(t), \tilde{x}^{\text{eff}}_{j'}(s)\right\}} \\
 & = \sum_{j}^{N} c^2_{j}\, \text{coth}\left(\frac{\beta \hbar \ff_{j}}{2} \right)
\text{cos}(\omega_j (t-s)) \, ,  
\end{split}
\end{equation}
\begin{equation}
\label{Force-7}
\begin{split}
 B_{\text{R}}(t,s) & = \eqexp{\sum_{j,j'}^{N}\frac{c_{j} c_{j'}}{2}
i\left[\tilde{x}^{\text{eff}}_{j}(t), \tilde{x}^{\text{eff}}_{j'}(s) \right]} \\
 & = \sum_{j}^{N} c^2_{j}\, \text{sin}(\omega_j (t-s)) \text{,}
\end{split}
\end{equation}
where we use $\{\cdot, \cdot \}$ to denote the anticommutator. These averages
over system-bath coupling operators characterize the fluctuations imposed
on the system via interaction with the bosonic bath. They
completely determine the impact of Gaussian fluctuations 
on the system under study \cite{Weiss_2012}. Hence, a  shift of 
the coupling operators $x_j$ to $x_j^{\text{eff}}(t)$ allows us to recover the
dynamics of the system in presence of an undriven bath in
thermal equilibrium. We note again, that this is only possible 
when the driving couples in a dipole-type manner to the individual 
bath oscillators. When the bath-driving would be parametric, the thermal 
fluctuations can be strongly altered. 

\subsection{Redefined effective Hamiltonian}
\label{sub:shift}

Exploiting the time-dependent shift of the position operator,  
we can now add Eq.\ \refq{Force-3} to the initial Hamiltonian and absorb the
effective time-dependent force into both the system and system-bath coupling
parts according to 
\begin{equation}
 \label{Force-8}
 \begin{split}
 H(t) & = H(t) - \langle H_{\rm SB} \rangle_{\rm B}(t) + \langle H_{\rm SB}
\rangle_{\rm B}(t) \\ 
 & = H_{\rm S}^{\rm eff}(t) + H_{\rm SB}^{\rm eff}(t) + H^{\rm eff}_{\rm B}(t)
 \end{split}
\end{equation}
with $H^{\rm eff}_{\rm B}(t)$ given in Eq.\ \refq{hbatheff}. As a first 
consequence the system-bath coupling operators are
shifted, as desired, and become time-dependent according to 
\begin{equation}
\label{effhamsb}
H_{\rm SB}^{\rm eff}(t) =
 -\frac{\sigma_{z}}{2} \sum_{j}^{N} c_j x_j^{\rm eff}(t) \, 
\end{equation}
with $x_j^{\rm eff}(t) = x_j - \eqexp{\tilde{x}_j(t)}$. In addition, the 
effective system Hamiltonian also becomes time-dependent as
\begin{equation}
\label{Force-9}
H_{\rm S}^{\rm eff}(t) = \frac{\Delta}{2} \sigma_x + \frac{F_{\text{eff}}(t)}{2}
\sigma_z
\text{.}
\end{equation}
In particular, the effective force $F_{\text{eff}}(t)$ introduces a
time-dependent asymmetry into the two-level system. This result leads to the
same conclusion as drawn from earlier findings 
\cite{GrabertPaper_2016_drivenbaths}. A dipole-type driving of bath modes 
yields an effective time-dependent force on the system. The force 
does not modify the fluctuational characteristics of the bath, but itself
depends on its prehistory, see Eq.\ \refq{Force-5}, i.e., on the full 
time interval $[t_a, t]$. In that sense, it may be denoted as a non-Markovian
force.

In the following, we work with the effective Hamiltonians of Eqs.\ 
\refq{effhamsb} and \refq{Force-9}, but will drop the superscript ``eff'' in
the symbol of the Hamiltonian from now on. 

\section{Adiabatic-Markovian master equation}

Equipped with the effective Hamiltonians of Eqs.\ \refq{effhamsb} and
\refq{Force-9}, we may now proceed to study the dissipative quantum dynamics.
In this work, we employ a master equation approach motivated by assuming a
weak system-bath coupling. In connection with the additional assumption of 
slow bath driving (the details are specified below), we can treat the influence
of the driven bath on the basis of a Born-Markov approximation, which
was previously used to investigate the dissipative Landau-Zener problem 
\cite{NalbachPaper_2013_am, NalbachPaper_2014_am}. 
A one-loop approximation of the self-energy then yields the quantum 
dynamics of the weakly damped and driven quantum two-level system in the form of 
 a simple Born-Markov approximated master equation \cite{NalbachDiss_1998, 
NalbachSolidsPaper_2002, NalbachPaper_2010_multiphonon, 
WuergerPaper_1997}. While its derviation follows different routes, the final 
result coincides with the standard Born-Markov quantum master equation, see , 
e.g., in Ref. \cite{Nitzan_2006}. Similar results could also 
be obtained employing resummation 
techniques within a path-integral framework 
\cite{Weiss1,Weiss2,Weiss3,Weiss4,Weiss_2012}. 

\subsection{Time-dependent rotation}
\label{sub:rot}

As a first step, we perform a time-dependent rotation into the momentary
eigenbasis of the effective system Hamiltonian \refq{Force-9} according to 
\begin{equation}
\label{AM-1}
 \bar{H}_{\rm S}(t) = R^{\dagger}(t) H_{\rm S}(t) R(t) = \frac{E(t)}{2} \tau_{x}
\text{,}
\end{equation}
with the momentary eigenenergies $E(t) = \sqrt{\Delta^{2} +
(F_{\text{eff}}(t))^{2}}$ and with $\tau_i$ denoting the Pauli matrices. The
rotation is generated by the operator 
$R(t) = \text{exp}\left[ i (\phi(t)/2)  \sigma_y \right]$ with 
the phase $\phi(t) = \text{arctan}\left[F_{\text{eff}}(t)/\Delta\right]$.
Rotation of the system-bath coupling Hamiltonian yields
\begin{equation}
\label{AM-2}
 \bar{H}_{\rm SB}(t) = -\left( \frac{u(t)}{2} \tau_z + \frac{v(t)}{2} \tau_x
\right) \sum_{j}^{N} c_{j} x_{j}^{\text{eff}}(t) \text{,}
\end{equation}
with the prefactors $u(t) = \cos \phi(t)$ and $v(t) =
\sin \phi(t)$. For later purposes, we also define a shifted system Hamiltonian
\begin{equation}
\label{AM-3}
 \bar{H}'_{\rm S}(t) = \bar{H}_{\rm S}(t) +
\frac{1}{2}\left(\frac{d\phi(t)}{dt}\right) \tau_y \, ,
\end{equation}
in which we take the time dependence of the phase into account.  

\subsection{Liouville space formulation}
\label{sub:liouville}

In order to evaluate the dynamics of the dissipative 
problem, we consider the total density matrix $W(t)$ of the 
system-plus-bath at time $t$ and 
make use of the Liouville-von Neumann equation
of motion 
\begin{equation}
\label{AM-4}
 \partial_t W(t) = -i[H(t),W(t)] \equiv \Ls(t) W(t) \text{,}
\end{equation}
with the time-dependent Liouvillian superoperator $\Ls(t) \, \cdot = -i[H(t),\,
\cdot\,]$ acting on operators in the product Hilbert space of system and
bath. The formal solution is given by 
\begin{equation}
\label{AM-5}
 W(t) = \mathcal{T}\,\text{exp}\left[\int_{t_{0}}^{t}ds \Ls(s)\right] W(t_0) =
\Us(t,t_0) W(t_0)  \text{,}
\end{equation}
with the time-evolution superoperator $\Us(t,t_0) =
\mathcal{T}\,\text{exp}\left[\int_{t_{0}}^{t}ds \Ls(s)\right]$ and 
$\mathcal{T}$ denoting the proper time-ordering operator.  Next, we assume
complete factorization of the initial total density matrix at coupling 
time $t_0$, such that $W(t_{0}) = \rho_{\rm S}(t_{0}) \otimes
\rho_{\rm B}(t_{0})$. Then, we
can average over the bath states to obtain the time-dependent reduced density
matrix of the system
\begin{equation}
\label{AM-6}
 \rho_{\rm S}(t) = \text{Tr}_{\rm B} \left[ \Us(t,t_0) W(t_0) \right] =
\Us_{\text{eff}}(t,t_0) \rho_{\rm S}(t_0)\text{.}
\end{equation}
Here, we have defined the effective time evolution superoperator
$\Us_{\text{eff}}(t,t_0) = \text{Tr}_{\rm B} \left[ \Us(t,t_0)
\rho_{\rm B}(t_0) \right] = \langle \Us(t,t_0) \rangle_{\text{B}}$ of the
reduced density matrix of the system. The time-evolution 
superoperator $\Us(t,t_0)$ can be expanded in a Dyson series and subsequent 
averaging over the bath modes then yields 
\cite{NalbachSolidsPaper_2002,NalbachPaper_2010_multiphonon} a 
similar expansion for the effective time-evolution superoperator
\begin{equation}
\label{AM-7}
\begin{split}
& \Us_{\rm eff}(t,t_0)  = \Us_{\rm S}(t,t_0) + \int_{t_0}^t ds\; \Us_{\rm
S}(t,s) \expe{\Ls_{\rm SB}(s) \Us_0(s,t_0)} \\
& + \int_{t_{0}}^t ds \int_{t_0}^s ds'\; \Us_{\rm S}(t,s) \expe{\Ls_{\rm SB}(s)
\Us_0(s,s') \Ls_{\rm SB}(s') \Us(s',t_0)} \text{,}
\end{split}
\end{equation}
where $\Us_{\rm 0}(t,t_0) = \Us_{\rm S}(t,t_0) \Us_{\rm B}(t,t_0)$ denotes the
uncoupled time-evolution with $\Us_{\rm S/B}(t,t_0)$ 
acting on the system or bath part, respectively. By combining Eqs.\ \refq{AM-6}
and \refq{AM-7}, we can recast the integral equation into the form 
of a master equation
\begin{equation}
\label{AM-8}
\begin{split}
\partial_t\rho &_{\rm S}(t) = \Ls_{\rm S}(t)\rho_{\rm S}(t) + \expe{\Ls_{\rm
SB}(t) \Us_0(t,t_0)}\rho_{\rm S}(t_{0}) \\ 
& + \int_{t_0}^t \hspace*{-2mm} ds\; \expe{\Ls_{\rm SB}(t) \Us_0(t,s) \Ls_{\rm
SB}(s) \Us(s,t_0)}\rho_{\rm S}(t_{0}) \text{.}
\end{split}
\end{equation}
Since the last term on the r.h.s.\ of this equation still contains the full
superoperator  $\Us(s,t_0)$, it is formally exact, but needs to be approximated
in order to allow for a practical solution. 

\subsection{Adiabatic-Markovian approximation}
\label{sub:am}

Due to the redefined system-bath coupling in Eq.\ \refq{Force-8}, the 
term of first order in the system-bath coupling, i.e., the second term on the
r.h.s.\ of Eq.\ \refq{AM-8}, vanishes. To see this, we note that 
$\Us_{\rm B}(t,t_0)\rho_{\rm B}(t_{0}) = \Us_{\rm B}(t,t_0)\Us_{\rm
B}(t_{0},t_{a})\rho_{\rm B}(t_{a}) = \Us_{\rm B}(t,t_a)\rho_{\rm B}^{\rm eq}$,
since we have defined that $\rho_{\rm B}(t_{a}) = \rho_{\rm B}^{\rm eq}$. In
this way, the first order term is proportional to $\eqexp{\tilde{x}_j^{\rm 
eff}(t)} = 0$.

The third term on the r.h.s.\ of Eq.\ \refq{AM-8} is approximated as
$\expe{\Ls_{\rm SB}(t)
\Us_0(t,s) \Ls_{\rm SB}(s) \Us(s,t_0)} \approx  \Ms(t,s) \Us_{\rm eff}(s,t_0)$
with the memory kernel $\Ms(t,s)=\eqexp{\Ls_{\rm SB}(t) \Us_0(t,s) \Ls_{\rm
SB}(s) \Us_{\rm B}(s,t_a)}$. This is 
the Born approximation which only keeps sequential one-phonon processes 
in a cumulant expansion diagrammatically representing a type of one-loop 
approximation scheme for system-bath correlations 
\cite{NalbachDiss_1998,WuergerPaper_1997,NalbachPaper_2010_multiphonon}. We 
note that we explicitly kept the uncoupled driven time evolution of the bath 
and have used the equilibrium average pertaining to 
$t_{a}$. In this way, the kernel is determined by Eqs.\ \refq{Force-6} and 
\refq{Force-7} and, thus, is essentially unchanged compared to the equilibrium 
situation apart from the particular time dependence of $u(t)$ and $v(t)$. 
Inserting both into Eq.\ \refq{AM-8} yields the Born-approximated quantum 
master equation 
\begin{equation}
\label{AM-9}
\partial_t\rho_{\rm S}(t) = \bar{\Ls}'_{\rm S}(t)\rho_{\rm S}(t) + \int_{t_0}^t
\hspace*{-2mm} ds\; \Ms(t,s)\rho_{\rm S}(s) \text{,}
\end{equation}
with the memory kernel $\Ms(t,s)$ given by
\begin{equation}
\label{AM-10}
\Ms(t,s) = \text{Tr}_{\rm B} \left[\bar{\Ls}_{\rm SB}(t) \Us_{0}(t,s)
\bar{\Ls}_{\rm SB}(s) \rho_{\rm B}(s) \right]\text{.}
\end{equation}
Here, we have restored the notation of the rotated Hamiltonians in
Eqs.\ \refq{AM-2} and \refq{AM-3}. 

In the next step, we assume a clear separation of time scales between the
dynamics associated with the system, the bath and the driving, such 
that the characteristic memory time $\tau_{\text{mem}}$ of the bath is much
shorter than both $\Delta^{-1}$ and the time scale associated with the driving
force. This adiabatic Markovian approximation builds on the observation 
that the memory kernel can then be assumed as short-lived, i.e.,
$\Ms(t-s) \ll 1$ for $t-s \gg \tau_{\text{mem}}$. As such, we approximate
the time evolution superoperator of the system as $\Us_{\rm S}(t,s) \approx
\text{exp}[\bar{\Ls}_{\rm S}(t)(t-s)]$ and the time-dependent
rotation parameters in $\bar{\Ls}_{\rm SB}(s)$ as $u(s) \approx
u(t)$ and $v(s) \approx v(t)$ \cite{NalbachPaper_2013_am, NalbachPaper_2014_am}. 
The additional time-dependence of $\bar{\Ls}_{\rm SB}(s)$, 
which enter via $x_{j}^{\rm eff}(t)$ in Eq.\ \refq{AM-2},
is left unchanged. In this way, we preserve 
the time-dependent shift of the coupling operators and retain the
exact equilibrium rates. Keeping this in mind, 
switching to the interaction picture and applying the
Markov approximation \cite{Nitzan_2006} according to
\begin{equation}
 \label{AM-11}
 \int_{t_0}^{t} \Ms(t,s) \rho_{\rm S}(s)~ds \approx \Ms_{\rm AM}(t)\rho_{\rm
S}(t) \, ,
\end{equation}
with 
$\Ms_{\rm AM}(t) = \int_{t_0}^{\infty} \Ms(t,s) e^{-\bar{\Ls}_{\rm S}(t)s}~ds$,
yields a Born-Markovian quantum master equation 
\begin{equation}
 \label{AM-12}
 \partial_t\rho_{\rm S}(t) = -i[\bar{H}'_{\rm S}, \rho_{\rm S}] -
\Gamma(t)\left[\rho_{\rm S}(t) - \rho^{\rm eq}_{\rm S}(t) \right] \, ,
\end{equation}
that depends parametrically on time $t$. 
Here, $\Gamma(t)$ is a momentary rate superoperator acting on both the reduced 
density matrix as well as the time-dependent pseudoequilibrium statistical
operator $\rho^{\rm eq}_{\rm S}(t)$.
In the case of weak system-bath coupling, this operator becomes $\rho^{\rm
eq}_{\rm S}(t) = \frac{1}{2}\left[\mathbb{1} - r_{x}^{\rm eq}(t) \tau_{x} 
\right]$ with $ r_{x}^{\rm eq}(t) = \text{tanh}\left[ \beta E(t)/2 \right]$ 
which is the result for a momentary thermal equilibrium. The rate coefficients 
in $\Gamma(t)$ can be obtained by explicitly writing down the 
kernel-superoperator
$\Ms_{\text{AM}}(t)$ as a matrix in Liouville space and evaluating its elements 
in Laplace space \cite{NalbachPaper_2010_multiphonon}. Furthermore, 
the imaginary parts of $\Ms_{\text{AM}}(t)$, which give rise to
frequency shifts, are neglected which is appropriate for weak system-bath
coupling. The secular approximation has been invoked as well.

\subsection{Generalized Bloch equations}
\label{sub:bloch}

On the basis of the adiabatic Born-Markovian approximated quantum
master equation \refq{AM-12} generalized Bloch equations can be
derived as usual \cite{Nitzan_2006}. We find for 
the expectation values $r_i(t) =-\langle\tau_i\rangle_t=
-\text{Tr}[\tau_i \rho_{\rm S}(t)]$ the equations of motion 
\begin{equation}
\label{AM-13} 
\begin{array}{lcl}
     \partial_t r_x(t) & = &  + \phi'(t) r_z(t) - \gamma_1(t)
[r_x(t)-r_x^{\rm eq}(t)] \, , \\
     \partial_t r_y(t) & = &  - \gamma_2(t) r_y(t) - E(t) r_z(t) \, ,\\
     \partial_t r_z(t) & = &  + E(t) r_y(t) - \gamma_2(t) r_z(t) - \phi'(t)
r_x(t)  \, ,
    \end{array}
\end{equation}
where the time derivative $\phi'(t) = d\phi(t)/dt$ of the mixing angle
is introduced via Eq.\ \refq{AM-3}. The time-dependent rate coefficients
follow as the time-dependent relaxation rate  
\begin{equation}
\label{AM-14} 
 \gamma_1(t) = \frac{1}{2} u^2(t)J(E(t))\text{coth}\left(\frac{\beta
E(t)}{2}\right) \, ,
\end{equation}
and the time-dependent dephasing rate 
\begin{equation}
\label{AM-15} 
 \gamma_2(t) = \frac{1}{2} \gamma_1(t) +
v^2(t)\left[J(\omega)\text{coth}\left(\frac{\beta
\omega}{2}\right)\right]\biggr|_{\omega\rightarrow 0} \text{.}
\end{equation}

\subsection{Generalized response}
\label{sub:resptheo}

To study the impact of the driven bath on the quantum two-level system, 
we consider the response function
\begin{equation}
\label{AM-17}
\begin{split}
R(t,t_{0})& = \text{Tr}_{\rm
S}\bigl\{i[\tilde{\sigma}_z(t,t_0),\sigma_z]
\rho_{\rm S}(t_{0})\bigr\} \\ 
& = \text{Tr}_{\rm S}\bigl\{\sigma_z \Us_{\rm eff}(t,t_{0}) i
[\sigma_z, \rho_{\rm S}(t_{0})] \bigr\}\text{.}
\end{split}
\end{equation}
A re-interpretation of this equation is convenient: 
it yields the expectation value of the operator 
$\sigma_z$ weighted by the operator $\Us_{\rm eff}(t,t_{0}) i
[\sigma_z, \rho_{\rm S}(t_{0})]$. In this sense, the latter operator may 
be identified as a different initial density matrix 
propagated by $\Us_{\rm eff}(t,t_{0})$, whose time-dependent elements
can be obtained using the Bloch equations \refq{AM-13}. 
The expectation value is then provided by the linear combination
$-[u(t)r_z(t) + v(t)r_x(t)]$. It is particularly convenient to study the 
frequency-dependent response function.
\begin{equation} \label{freqres}
R(\ff)= \int dt \, e^{i \ff t} \, R(t,t_{0})\, .
\end{equation}

\section{Bath driving pulses}

In this work, we consider two particular bath-driving shapes: 
a Dirac $\delta$-pulse as well as a Gaussian driving-pulse. From this point 
onwards, we set $t_{0} = 0$ for simplicity and fix $t_a$ and $t_g$ 
separately. First, we consider a Dirac $\delta$-shaped driving 
pulse acting at $t=t_a$ with area $\Delta^{-1}$
\begin{equation}
\label{Res-1}
 F^{\delta}(t) = \Delta^{-1} \delta(t - t_{a}) \text{,}
\end{equation} 
which generates the effective force 
\begin{equation}
\label{Res-2}
F_{\text{eff}}^{\delta}(t) = -\frac{1}{\Delta \pi} \int_{0}^{\infty}d\ff \bar{J}(\ff)
\sin\omega (t - t_a)\text{.}
\end{equation}

As a second case, we consider a Gaussian-shaped pulse with area $\Delta^{-1}$
\begin{equation}
\label{Res-3}
 F^{g}(t) = \frac{\Delta^{-1}}{\sqrt{2 \pi} \sigma}
 e^{-\frac{(t - t_{g})^{2}}{2\sigma^{2}}} \text{,}
\end{equation} 
centered at $t=t_g$ and with a width $\sigma$. It generates a force 
\begin{equation}
\label{Res-4}
 F_{\text{eff}}^{g}(t) = \text{Im}\left[\frac{1}{2\pi \Delta} 
 \int_{0}^{\infty}d\ff \bar{J}(\ff) e^{- \frac{\ff^2 \sigma^{2}}{2} - i\ff (t - 
t_{g})} \text{erfc}(\zeta_{t}) \right]\text{,}
\end{equation} 
with $\zeta_{t} = (i\ff\sigma^{2} - t + t_{g})/\sqrt{2\sigma^{2}}$ and
the complementary error-function $\text{erfc}(z)$. For the derivation
of Eq.\ \refq{Res-4}, we have assumed the Gaussian at $t_a$ to be sufficiently 
small and far away from the center $t_g$, such that the whole Gaussian
is eventually taken into account during the integration in Eq.\ \refq{Force-5}.

To fully characterize the effective force, we also need knowledge about the
additional spectral density $\bar{J}(\ff)$ defined in Eq.\ \refq{Force-4}.
As shown in Ref.\ \cite{GrabertPaper_2016_drivenbaths} for two particular
examples of applications, a simple proportionality $\bar{J}(\ff) \propto
J(\ff)$ can be found. This result stems from a model of a polar 
environmental solvent and involves linear susceptibilities for the response to 
emerging electrical reaction fields. Explicitly calculating the 
additional field-contributions from bath driving allows one to derive 
aforementioned proportionality. We will make use of this result here and choose 
the proportionality factor individually, see below. Furthermore, it 
should be noted that the magnitude of the force in Eqs.\ \refq{Res-1} and 
\refq{Res-3} was absorbed into the definition of $\bar{J}(\ff)$ such that the 
prefactor of the latter ultimately determines the strength of the external 
driving.

In passing, we also note that both bath-driving pulse-shapes 
eventually subject the TLS to effective driving pulses of finite duration. The 
impact of finite pulses on the transition probability has already been 
investigated for pulses with various shapes \cite{Vitanov_2004,Vitanov_2013,Conover2011}. 
In our case, a preliminary investigation of the excitation probability for the 
cases and parameters considered (not shown) leads to an oscillatory behaviour 
reminiscient of the $\text{sin}^{2}$-dependence known from the Rabi-formula for 
rectangular pulse-shapes \cite[Eq.\ (2)]{Vitanov_2013}. 
A detailed analysis may be subject of future works.

\section{Dynamics in a driven Ohmic bath}
\label{sub:ohmian}

First, we study the dynamical properties of the quantum two-level system 
in a bath with a generic Ohmic spectral density 
\begin{equation}
 \label{Res-5}
 J(\omega) = \frac{\eta \omega}{\omega_c} e^{-\omega/\omega_{c}}
\end{equation}
with an exponential cut-off, where $\omega_c$ is the cut-off frequency. In
addition, we set $\bar{J}(\ff) = (\bar{\eta}/\eta) J(\ff)$. For simplicity,
we evaluate the dynamics at zero temperature.

\subsection{Dirac pulse}

The effective force in Eq.\ \refq{Res-2} for the Dirac $\delta$-pulse can be
obtained analytically as 
\begin{equation}
 \label{Res-6}
 F_{\text{eff}}^{\delta}(t) = -\frac{2 \ff_{c} \bar{\eta}}{\pi \Delta}
\frac{\ff_{c} (t - t_a)}{[1 + \ff_{c}^{2} (t - t_a)^{2}]^{2}}\text{.}
\end{equation}
\begin{figure}[t!]
\includegraphics[width=6.5cm]{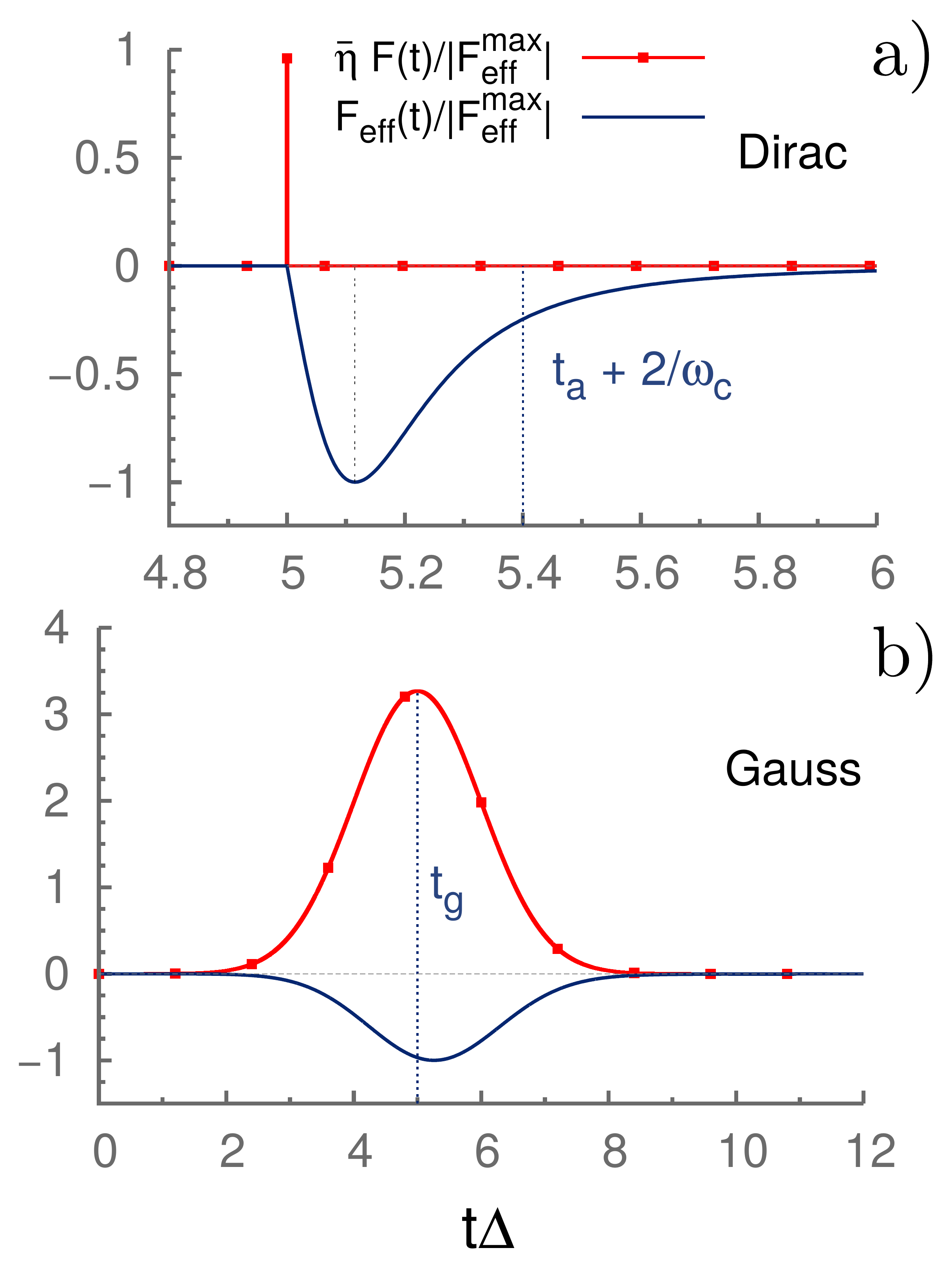}
\caption{\label{fig-2} Normalized effective force $F_{\text{eff}}(t)$ 
(blue solid line) and direct driving force $\bar{\eta} F(t)$ 
(red line with squares) for an Ohmic bath driven by a Dirac $\delta$-pulse 
(a) and a Gaussian pulse (b). The bath is characterized by the parameters $\eta 
= 0.05\Delta$ and $\omega_c = 5\Delta$. The Dirac pulse (a)
occurs at $t_a = 5\Delta^{-1}$ with interaction strength 
$\bar{\eta} = 2\Delta$. The Gaussian pulse (b) is centered
at $t_g = 5\Delta^{-1}$ and starts at $t_a = 0$ with interaction
strength $\bar{\eta} = 6\Delta$ and width $\sigma = \Delta^{-1}$.
Both quantities are normalized with respect to the 
maximum of the effective force to allow for a comparison of relative 
strengths. Notice that the height of the Dirac pulse has also been chosen to 
correspond to its effective strength as well.
}
\end{figure}
It is depicted in Fig.\ \ref{fig-2} a) for the set of parameters as indicated.
In addition, the same figure shows the direct driving force
$\bar{\eta}F(t)$ for comparison. The retardation and the
decay on the characteristic time scale $1/\ff_{c}$ are apparent. 

\begin{figure}[th]
\includegraphics[width=6.5cm]{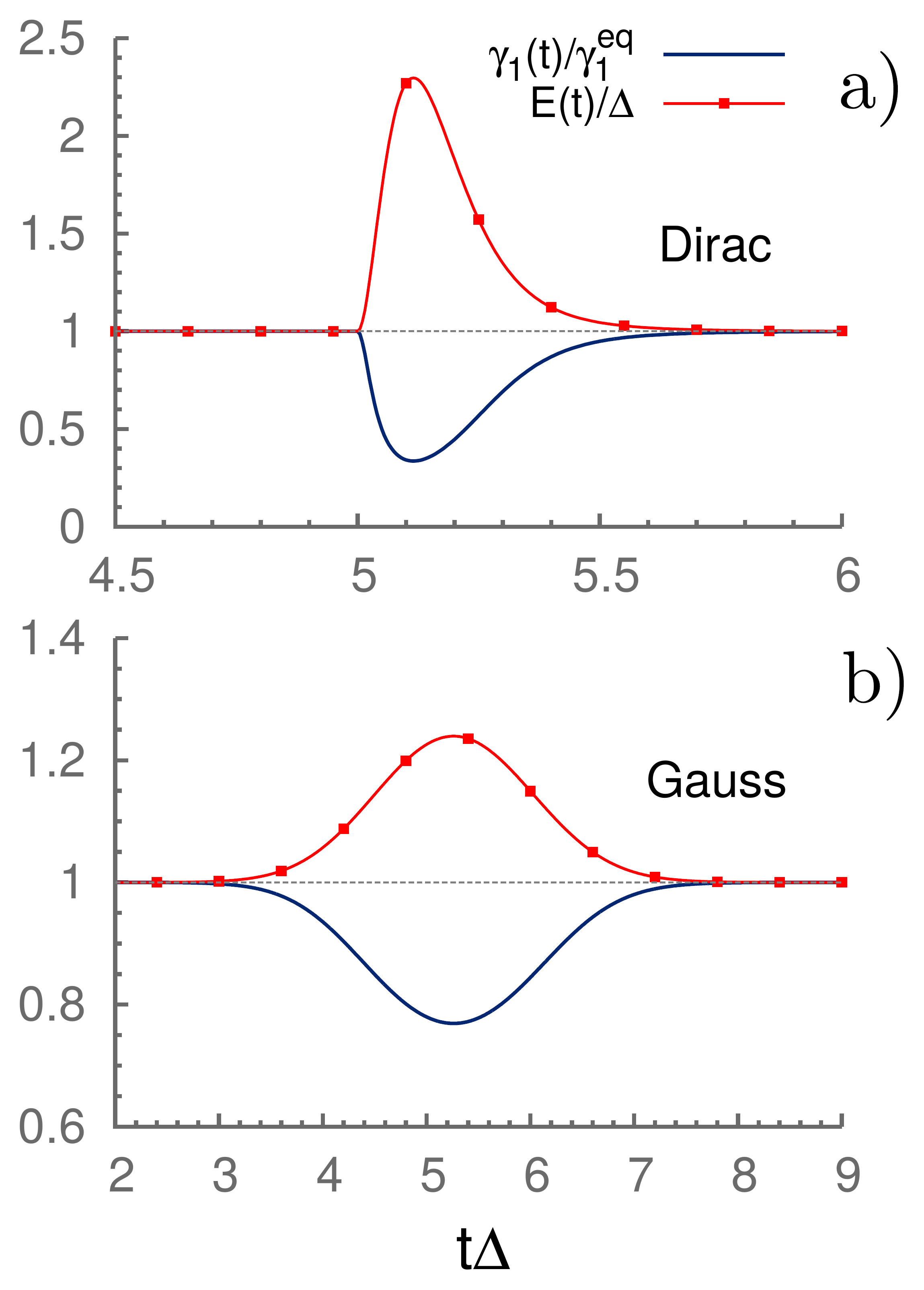}
\caption{\label{fig-3} Time-dependent relaxation rate (blue solid line)
and momentary energy (red line with squares) for the Ohmic bath driven by a 
Dirac (a) and a Gaussian (b) pulse at zero temperature. The parameters are the 
same as in Fig. \ref{fig-2}. The normalization has been chosen with respect to  
the undriven (equilibrium) relaxation rate $\gamma_1^{\rm eq} = 
J(\Delta)/2$ and the bare, undriven system energy 
scale $\Delta$.
}
\end{figure}
%
The ratio of the time-dependent relaxation rate of Eq.\ \refq{AM-14} and
its equilibrium value is shown in Fig.\ \ref{fig-3} a). Driving leads to a
visible reduction near the onset of the driving.
\begin{figure}[t]
\includegraphics[width=6.5cm]{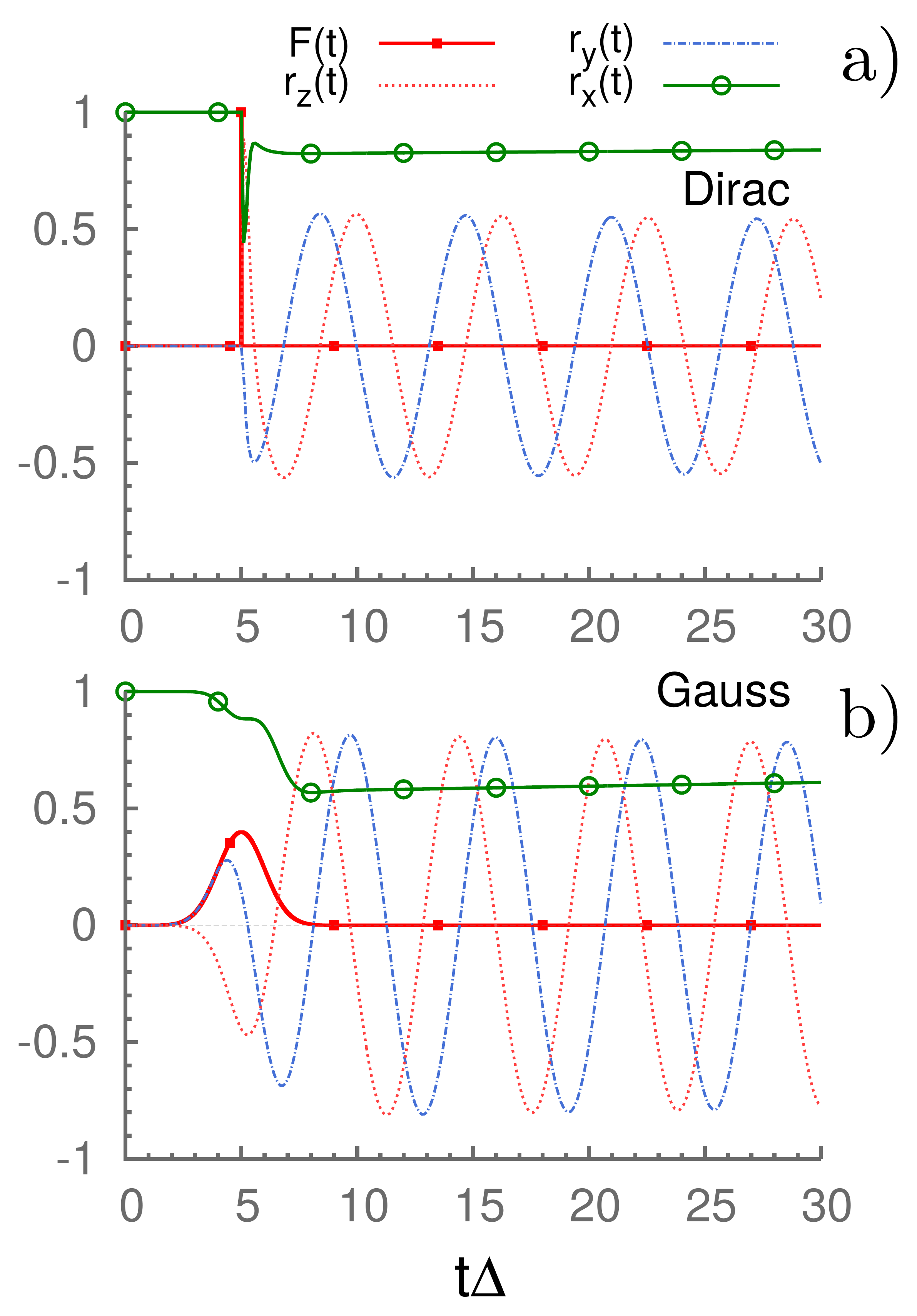}
\caption{\label{fig-4} Dynamics of the expectation 
values $r_i(t) = -\langle \tau_i\rangle_t$ (green solid line with 
empty circles, dashed-dotted blue line and dotted red line) of the two-level 
system in an Ohmic bath at zero temperature, driven by a Dirac (a) and a 
Gaussian (b) pulse. The (dimensionless) external driving force $F(t)$ is 
also shown (red solid line with squares) as a comparison. The dynamics are 
generated by the forces and rates shown in Figs.\ \ref{fig-2} and \ref{fig-3}, 
with parameters being the same as in Fig. \ref{fig-2}. The system is set to be 
initially in equilibrium, i.e., in the ground state.}
\end{figure}
%
The resulting time evolution of the components $r_i(t)=-\langle \tau_i\rangle_t$
of the reduced density matrix are shown in Fig.\ \ref{fig-4} a). The system 
dynamics broadly follows the effective force profile, with the momentary 
population difference $r_x(t)$ changing rapidly near the Dirac pulse. This 
indicates an excitation of the ground state on time scales
determined by the effective force. In addition, an equally 
abrupt emergence of coherences is also visible. For longer times, the effective
force vanishes and $r_x(t)$ decays back exponentially with a rate constant
given by its equilibrium rate. However, we should keep in mind that the
characteristic decay on a time scale 
$1/\ff_{c}$ means that a comparably rapidly changing force is present. Then, 
the adiabatic-Markovian approximation may be problematic in this particular
case.

\subsection{Gaussian pulse}
The effective force generated by a Gaussian pulse acting on the bath, is
shown in Fig.\ \ref{fig-2} b) and the resulting relaxation rate in Fig.\ \ref{fig-3}
b).  The dynamics of the elements  $r_i(t)=-\langle
\tau_i\rangle_t$ of the reduced density matrix is shown in Fig.\ \ref{fig-4}
b). The effective force follows the perturbation closely but also shows a clear
retardation and fast decay as soon as the external perturbation effectively terminates.
The rate and dynamics of the density matrix behave roughly in the same way 
as in the Dirac case, where the time-dependent rate is reduced as long as the 
effective force is active and the pulse leads to evident excitation of the TLS 
and subsequent decay with the equilibrium rate for longer times. Visible 
differences only occur when the Gaussian is still active. Instead of fast 
excitation, a plateau-like behaviour and smooth emergence of coherent 
superpositions can be observed. In contrast to the Dirac case,
the emerging effective force is also smaller than the initial perturbation.

\section{Dynamics in a driven Lorentzian bath}
\label{sub:lorentzian}
Another interesting class of bath spectral densities describes structured
baths. A structured bath may be characterized by a Lorentzian spectral density 
\begin{equation}
 \label{Res-7}
 J(\omega) = \kappa \frac{\Gamma \Omega^{2} \omega}{
(\omega^2 - \Omega^{2})^{2} + (\Gamma \omega)^{2}}\text{,}
\end{equation}
which has a Lorentzian peak centered at a given frequency $\Omega$ with a width
$\Gamma$. This additional peak may be associated with a distinct bath mode
\cite{GrabertPaper_2016_drivenbaths} and may give rise to interesting resonance
effects. Instead of the structureless Ohmic spectral density 
of Eq.\ \refq{Res-5}, the Lorentzian peak introduces a pronounced oscillatory 
component into the frequency response of the bath. This may be understood in 
terms of a convenient mapping of the Lorentzian bath onto a single 
harmonic oscillator with frequency $\Omega$ which itself is coupled to a
structureless Ohmic bath \cite{ThorwartPaper_2004_structured,GargPaper_1985}. 
For the case considered in this work, the coupling of the system to the
single mode is given by $g = \sqrt{\kappa \Omega/8}$
and the coupling of the mode to the Ohmic bath is given by $h = \Gamma/(2\pi
\Omega)$. Here, we will calculate the dynamics in the original 
system and use aforementioned mapping for the analysis of the 
frequency-dependent response in section \ref{sub:freq_resp}. As before, we 
set $\bar{J}(\ff) = (\bar{\kappa}/\kappa) J(\ff)$ and evaluate the dynamics at 
zero temperature.

\begin{figure}[t!]
\includegraphics[width=6.5cm]{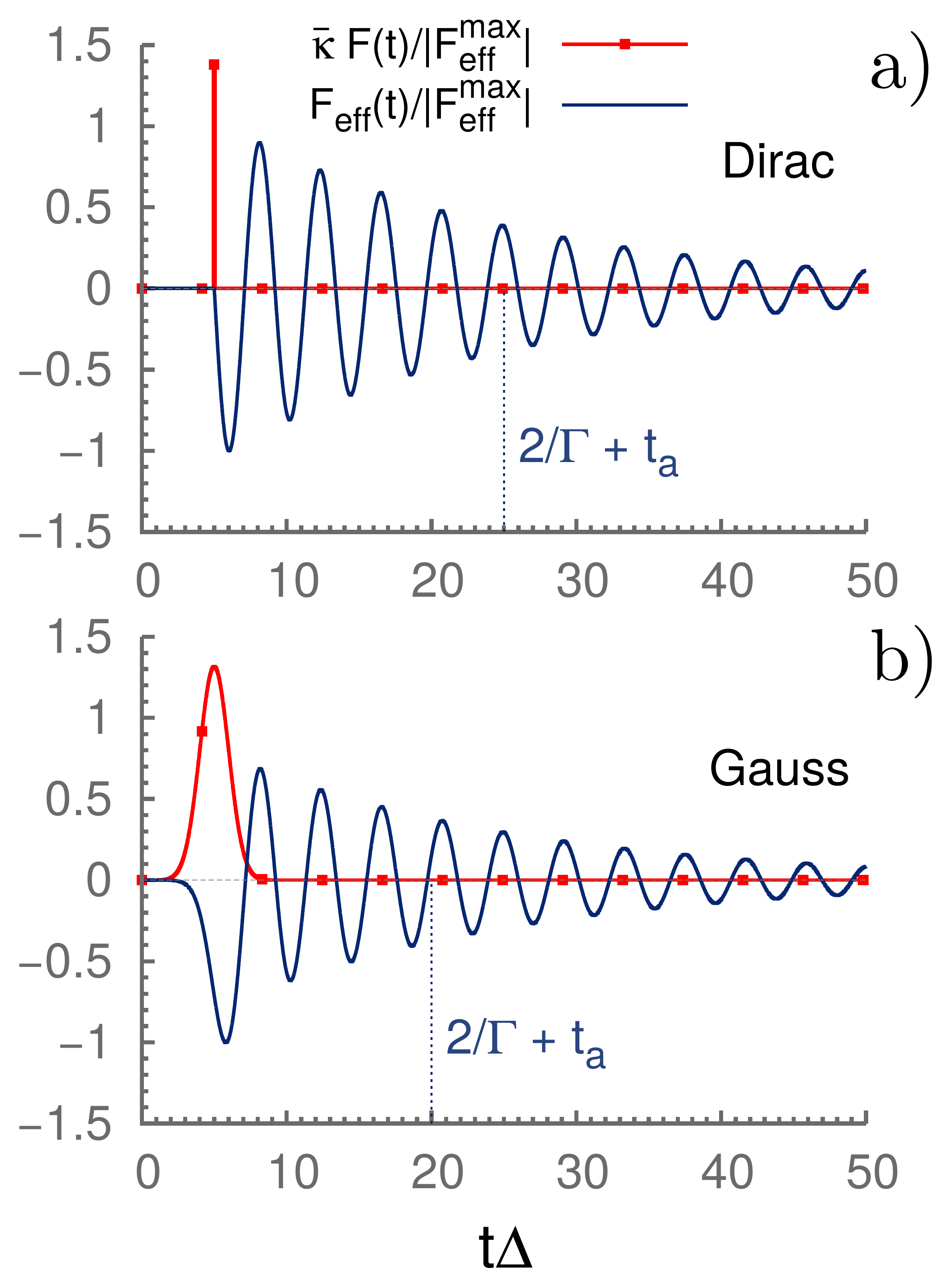}
\caption{\label{fig-5} Normalized effective force (blue solid line) and
direct driving force $\bar{\kappa}F(t)$ (red line with squares) for a driven 
Lorentzian bath with a Dirac (a) and a Gaussian (b) driving pulse. 
The bath is characterized by the parameters $\kappa = 0.05\Delta$,
$\Omega = 1.5\Delta$ and $\Gamma = 0.1\Delta$. The Dirac pulse (a)
occurs at $t_a = 5\Delta^{-1}$ with interaction strength 
$\bar{\kappa} = 2\Delta$. The Gaussian pulse (b) is centered
at $t_g = 5\Delta^{-1}$ and starts at $t_a = 0$ with interaction
strength $\bar{\kappa} = 6\Delta$ and width $\sigma = \Delta^{-1}$.
Both quantities are normalized with respect to the 
maximum of the effective force to allow for a comparison of relative 
strengths. Notice that the height of the Dirac pulse has also been chosen to 
correspond to its effective strength as well.
}
\end{figure}
\subsection{Dirac pulse}
The effective driving force and the direct bath driving force for the Lorentzian
bath are shown in Fig.\ \ref{fig-5} a) for a Dirac pulse. An oscillatory decay
emerges which can be fitted by a function $f(t) = - e^{-\Gamma t/2}\sin \Omega
t$, which originates from the Lorentzian peak in the spectral density. 
\begin{figure}[t]
\includegraphics[width=6.5cm]{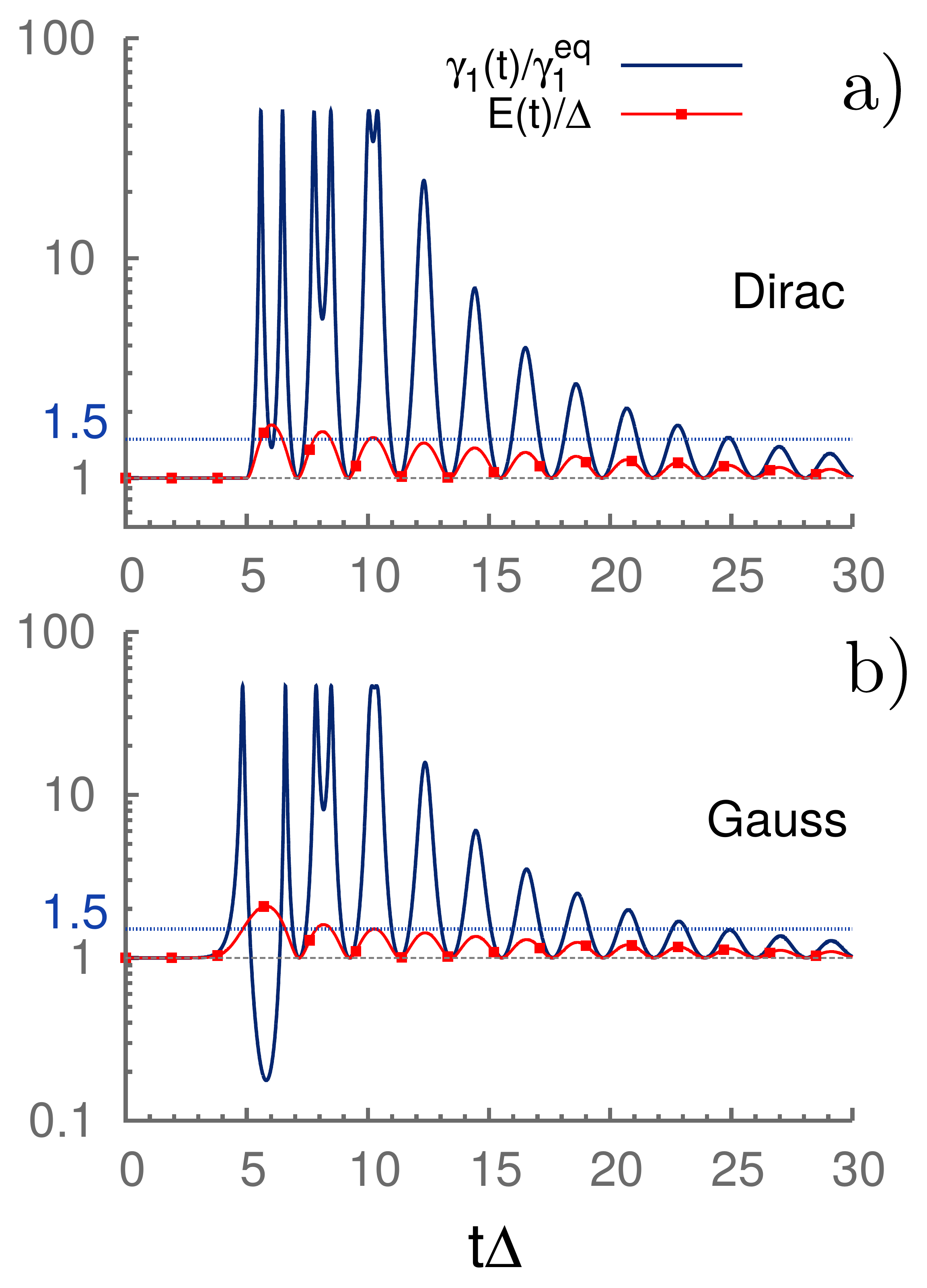}
\caption{\label{fig-6} Time-dependent relaxation rates (blue solid line)
and momentary energy (red line with squares) for a Lorentzian bath driven by a 
Dirac (a) or a Gaussian (b) pulse at zero temperature. The parameters are the 
same as in Fig. \ref{fig-5}. The normalization has been chosen with respect to  
the undriven (equilibrium) relaxation rate $\gamma_1^{\rm eq} = 
J(\Delta)/2$ and the bare, undriven system 
energy scale $\Delta$.
}
\end{figure}
The time-dependent zero-temperature rate is shown in Fig.\ \ref{fig-6} a) and
behaves in a somewhat more peculiar way,  with a strong alternating 
enhancement and suppression appearing as pronounced peaks. The rate peaks show a
characteristic splitting whenever $E(t) \geq \Omega$. It vanishes as soon 
as the momentary energy becomes smaller. The splitting is a signature that 
enough energy for the excitation of the harmonic mode at $\Omega$ is available 
which can then be used as a secondary relaxation pathway. 
\begin{figure}[t]
\includegraphics[width=6.5cm]{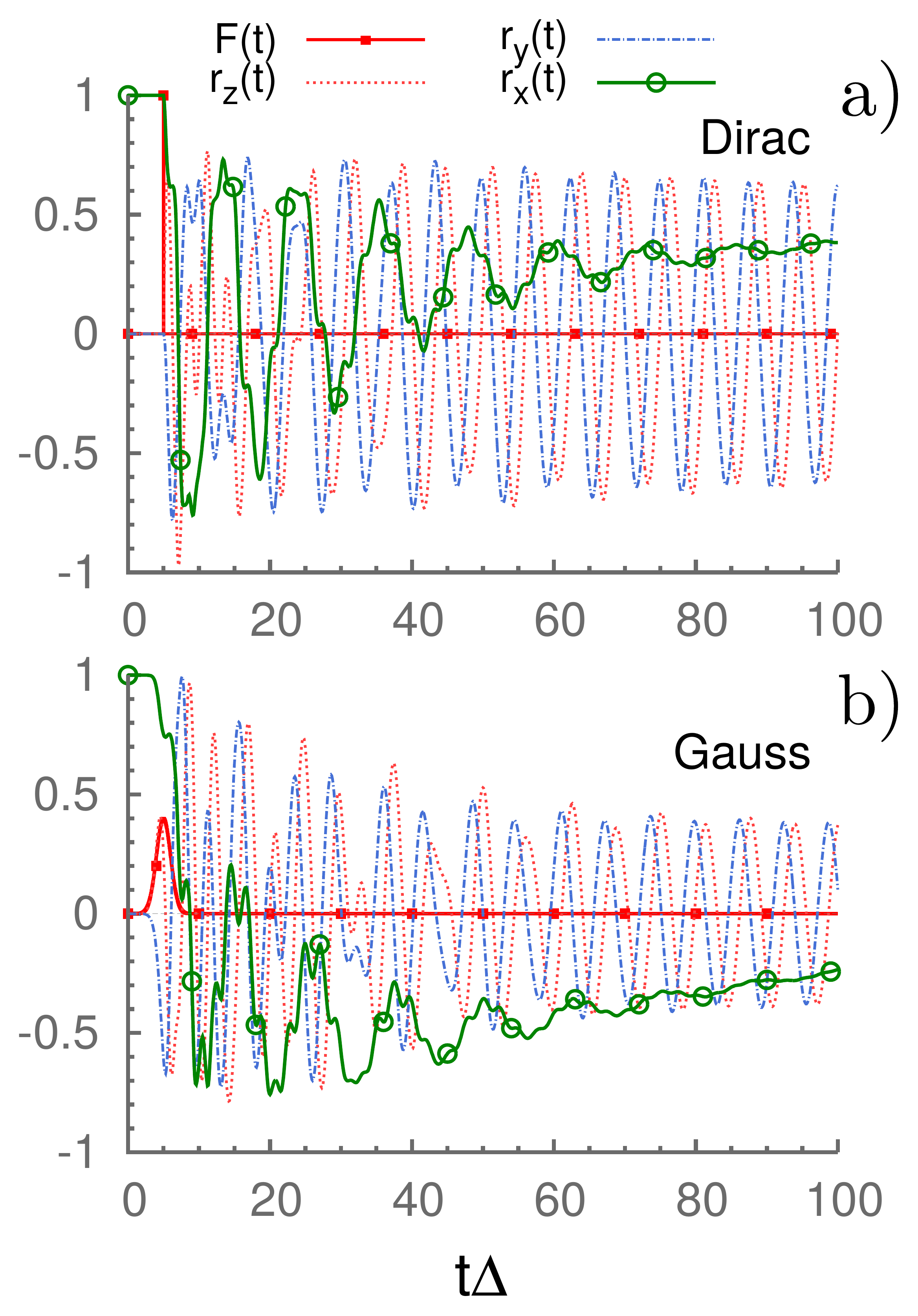}
\caption{\label{fig-7} Dynamics of the expectation 
values $r_i(t) = -\langle \tau_i\rangle_t$ (green solid line with 
empty circles, dashed-dotted blue line and dotted red line) of the two-level 
system in a Lorentzian bath at zero temperature, driven by a Dirac (a) and a 
Gaussian (b) pulse. The (dimensionless) external driving force $F(t)$ is 
also shown (red solid line with squares) as a comparison. The dynamics are 
generated by the forces and rates shown in Figs.\ \ref{fig-5} and \ref{fig-6}, 
with parameters being the same as in Fig.\ \ref{fig-5}. The system is set to be 
initially in equilibrium, i.e., in the ground state.}
\end{figure}
In terms of the dynamics of the density matrix components  $r_i(t)$ shown in 
Fig.\ \ref{fig-7} a), the interaction with the strongly pronounced harmonic
mode is visible via rapid oscillations with diverse 
frequency components both in the population difference as well as in the
coherences. The rapid fluctuations are damped with increasing time leading to
undriven exponential decay when the effective force has vanished. 

\subsection{Gaussian pulse}
The effective force generated by a Lorentzian bath driven by a
Gaussian pulse is shown in Fig.\ \ref{fig-5} b). Its behaviour 
is similar to the case of a Dirac pulse, but slight differences at short 
times occur due to the non-zero extent of the Gaussian pulse in time. 
The relaxation rate and the dynamics for the case of a Gaussian bath-driving 
pulse are shown in Fig.\ \ref{fig-6} b) and \ref{fig-7} b), respectively. 
Again, they show a qualitatively similar behaviour as in the previous Dirac 
case, with only minor differences arising when the Gaussian is still active, 
i.e., within a few widths of $t_g$.

\section{Frequency-dependent response}
\label{sub:freq_resp}
In this section, we evaluate the response function of a quantum two-level 
system to a driven harmonic bath. For a structureless driven Ohmic bath, it may 
be expected that the frequency-dependent response is only quantitatively 
different from the case when the bath is undriven. The situation is different 
for a structured Lorentzian bath, since additional resonances may be expected 
due to the interplay of the distinct environmental mode with the central 
system. 
  
\subsection{Driven Ohmic bath}
\label{sub:respresOhm}

\begin{figure}[t]
\centering
\includegraphics[width=8.5cm]{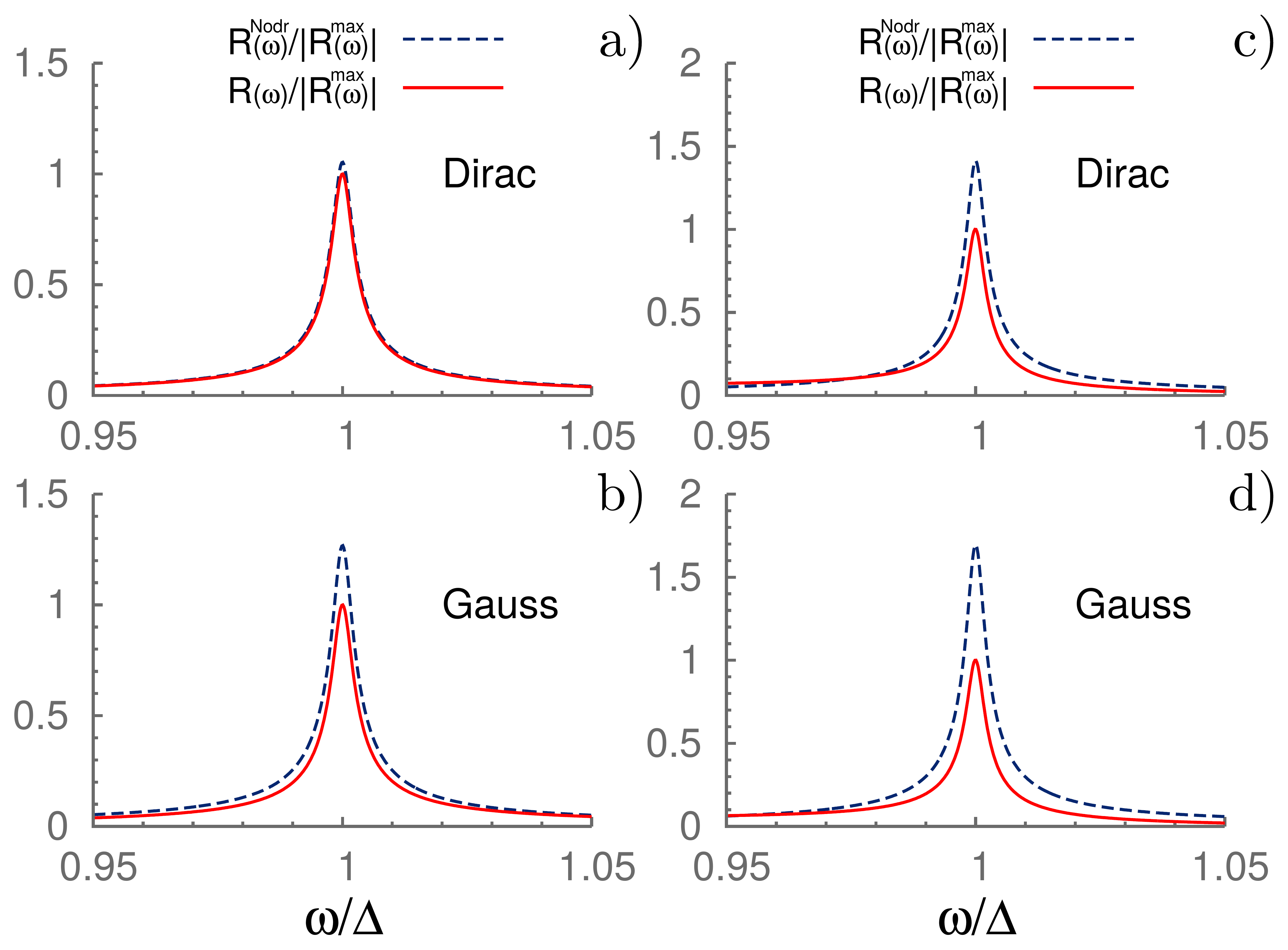}
\caption{\label{fig-8} Frequency-dependent response (red solid line) of 
a quantum two-level system to an Ohmic (a,b) and a Lorentzian (c,d) bath at 
zero temperature driven by a Dirac (a,c) or a Gaussian (b,d) pulse. For 
comparison, the response to an undriven bath is also shown (blue dashed 
line). The parameters used are given below Fig.\ \ref{fig-2} for the Ohmic 
bath and below Fig.\ \ref{fig-5} for the Lorentzian bath. Both quantities 
have been normalized with respect to the maximum of the driven frequency 
response to allow for a comparison of relative strengths.}
\end{figure}

The frequency-dependent system response of Eq.\ \refq{freqres} for the case of 
a driven Ohmic bath is shown in Fig.\ \ref{fig-8} a) and b) 
for both driving pulse shapes in comparison to the 
response without bath driving. In general,  
Lorentzian-shaped response characteristics result, with the maximum centered 
at $\ff = \Delta$. In both cases, bath driving leads to a reduction of the 
central peak height, which 
indicates that driving of an Ohmic bath leads to less effective direct driving. 
For the Gaussian pulse, this effect is more pronounced, since the peak 
is reduced more strongly by about $30\%$ in comparison to the undriven case.

\begin{figure}[t]
\centering
\includegraphics[width=8.5cm]{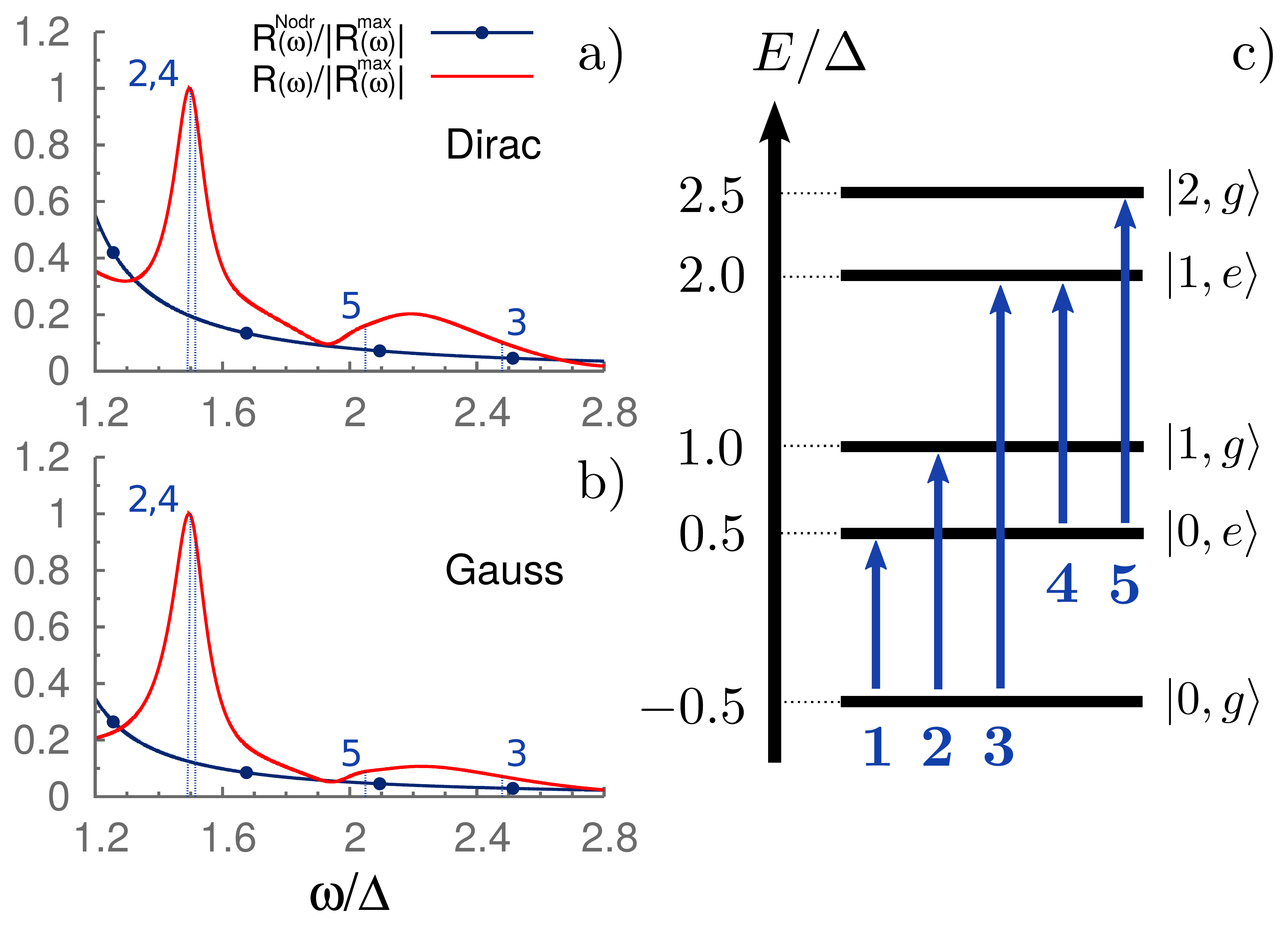}
\caption{\label{fig-9} Frequency-dependent response (red solid line) to a 
Lorentzian bath at zero temperature driven by a Dirac (a) or a Gaussian (b) 
pulse away from the fundamental frequency $\Delta$.  For comparison, the 
response to an undriven bath is also shown (blue line with circles). 
The parameters used are given below Fig.\ \ref{fig-5}. Both quantities 
have been normalized with respect to the maximum of the driven frequency response to allow 
for a comparison of relative strengths. The emerging peaks correspond well to 
energy gaps (blue dotted lines) obtained numerically from the 
Hamiltonian in Eq.\ \refq{Res-8} with $g = \sqrt{\kappa \Omega/8} \approx 
0.1\Delta$ ($\Omega = 1.5\Delta$). The level diagram for $g = 0$ (c) 
shows the corresponding transitions (blue arrows).
}
\end{figure}

\subsection{Driven Lorentzian bath}
\label{sub:respresLor}
The picture is more involved in the case of a Lorentzian bath, where the
resonant interaction of the two-level system with the driven pronounced bath
mode at frequency $\Omega$ can become possible.
In Fig.\ \ref{fig-8} c) and d), the frequency-dependent response close to theF
main frequency $\omega \approx \Delta$ is shown. As in the Ohmic case,
the response at the main frequency is reduced when the bath-driving is included.
In addition, further resonant response peaks arise which are shown in 
Fig.\ \ref{fig-9} a) and b). These additional resonant peaks can be 
understood when the mapping outlined in section \ref{sub:lorentzian} is used. 
Thus, we consider the TLS coupled to a structured bath by using the 
equivalent situation when a TLS-plus-harmonic-oscillator is coupled to a 
structureless bath. The Hamiltonian of this two-level system coupled to a 
single harmonic oscillator with frequency $\Omega$ and coupling strength $g$, 
is \cite{ThorwartPaper_2004_structured}
\begin{equation}
\label{Res-8}
 H_{\text{TLS-HO}} = \frac{\Delta}{2}\sigma_x - g\sigma_{z}(B^{\dagger} + B) +
\Omega B^{\dagger}B \, .
\end{equation}
Here, $B$/$B^{\dagger}$ are the annihilation/creation operators of the harmonic
oscillator. The energy level scheme of the combined TLS-plus-oscillator 
system is shown in Fig.\ \ref{fig-9} c) for vanishing 
coupling $g$. The corresponding transition frequencies for finite $g$ obtained
from numerical diagonalization
are marked in Fig.\  \ref{fig-9} a) and b) by blue dotted lines and correspond
well with the additional 
peaks obtained. Notable in this case is the existence of transitions
from the excited TLS state (transitions 4 and 5) and both the lack 
of an observable shift in the two-level transition peak as well 
as the lack of observable level splitting between the transitions 2 and 4. 

\section{Conclusions}

When an open quantum system is driven by an external time-dependent field, it 
is often unavoidable in principle that the driving also couples to 
the environment. Usually, this effect is neglected in the theoretical 
description. In a sense, a special case of a driven bath is given by a pumped 
optical resonator in which an atom is placed. Our approach addressed 
a more general case by considering a continuous distribution of 
bath modes which can be driven.  

Subsequently, we have shown that bath driving which couples 
linearly to the displacements of the bath oscillators (dipole-type driving) 
generates an additional time-dependent force for the central system. This 
effective force is retarded and depends on the entire time range from its onset 
to the momentary time as well as the spectral characteristics of the bath. We
investigated this effect for the case of the spin-boson model in the weak 
system-bath coupling regime. In order to illustrate the emerging dynamics, we 
generalized a Born-Markovian quantum master equation approach in which 
a certain class of terms in the Liouvillian superoperator are summed up after a 
linearization in the system-bath coupling, while the effective force was 
assumed to be slow. The time-dependent bath-induced force then leads to 
time-dependent rate coefficients in the quantum master equation 
which can be solved numerically.

To be specific, we considered two types of bath spectral densities, 
the standard Ohmic bath and the structured Ohmic bath in which a distinct bath 
mode has a peaked spectral weight. Furthermore, we calculated the bath-induced 
force for two types of bath-driving, a Dirac delta-shaped pulse and a 
Gaussian-shaped pulse. We found that the response of the central system 
including the bath-induced force is significantly modified.  
For the unstructured Ohmic bath, the resonant response of the quantum 
two-level system is effectively reduced when bath driving is included. 
Interestingly enough, a qualitatively different response arises 
when a structured Ohmic bath with a Lorentzian peak in the environmental 
spectral density is considered. Then, additional resonant peaks appear in the 
response of the system when the external drive excites the pronounced bath 
mode.

Since driven dissipative quantum systems are ubiquitous, the effect 
described in the present work should be considered in an accurate theoretical 
description of the time-dependent response and may
provide a basis for new, elaborate driving schemes. 

\acknowledgments
We acknowledge financial support of ``The Hamburg Centre for Ultrafast
Imaging (CUI)'' within the German Excellence Initiative supported by the
Deutsche Forschungsgemeinschaft.

\bibliographystyle{apsrev4-1}

\end{document}